# ADDING A NEW SITE IN AN EXISTING ORACLE MULTIMASTER REPLICATION WITHOUT QUIESCING THE REPLICATION

Hakik Paci[1], Elinda Kajo[2], Igli Tafa[3] and Aleksander Xhuvani[4]

[1]Department of Computer Engineering, Polytechnic University of Tirana, Albania
hpaci@fti.edu.al
[2]Department of Computer Engineering, Polytechnic University of Tirana, Albania
ekajo@fti.edu.al
[3]Department of Computer Engineering, Polytechnic University of Tirana, Albania
itafa@fti.edu.al
[4]Department of Computer Engineering, Polytechnic University of Tirana, Albania
axhuvani@fti.edu.al

*ABSTRACT*

*This paper presents a new solution, which adds a new database server on an existing Oracle Multimaster Data replication system with Online Instantiation method. During this time the system is down, because we cannot execute DML statements on replication objects but we can only make queries. The time for adding the new database server depends on the number of objects, on the replication group and on the network conditions. We propose to add a new layer between replication objects and the database sessions, which contain DML statements. The layer eliminates the system down time exploiting our developed packages. The packages will be active only during the addition of a new site process and will modify all DML statements and queries based on replication objects.*

## 1. INTRODUCTION

Replication is the process of copying and maintaining database objects, such as tables, in distributed database system. Changes applied at one site are stored locally in a stage before being forwarded and applied at each of the remote locations.

Some of the most common reasons for using replication are described as follows:

- Performance and Availability. Users are accessing data from a local database instead of a remote database, so replication provides fast access to shared data. Some users can access one database server while other users can access different database servers which all provide the same data, thereby reducing the workload of all servers [1].

- Network Load Reduction. Replication can be used to distribute data over multiple regional locations. At that point, applications can access various regional servers instead of accessing one central server. This configuration can reduce network load dramatically [1].

Another advantage of using replication is the fact that we can also use offline applications like mobile databases to be updated. Also we can balance the load distribution to several nodes.

During adding a new master site the system is down, because we cannot execute DML statements on replication objects but we can only make queries. The time for adding the new database server depends on the number of objects, on the replication group and on the network conditions. We propose to add a new layer between replication objects and the database sessions, which contain DML statements. The layer eliminates the system down time exploiting our developed packages.

## 2. REPLICATION

The following sections explain what is the replication and the basic components of a replication system, including replication objects, replication groups and replication sites.

### 2.1. What is replication?

Three main requirements of database replication are the performance, the availability and the consistency of data. These requirements are in conflict with each other because a change for the benefit of one of the criterion implies a change (minimization) at the expense of the other criteria.

Ensuring consistency is the main problem with replication of data in every particular copy of the database. Making sure that conflict operations (e.g. a read and write operation at the same data acting concurrently) are done in the same order at every replica ensures us that replicas are consistent. According to the consistency there are two traditional model classifications, strong and weak consistency models [2]. More clearly, strong consistency models ensure that the content of all of the replicas is the same as before any transaction. In an unreliable network like the internet, when we deal with large amount of replicas sometimes it is impossible to use strong consistency because of the high latency which affects in not having synchronous replicas in appropriate time. In consequence the strong consistency model is only suitable for systems with few replicas. It is also possible to use weak consistency as an alternative. In this consistency model, to ensure transactions, it is not necessary that all of the replicas have the same content. It is only needed to ensure that the replicas sometime converge to a consistent state, defined by synchronization points, which is not bounded to a specific period of time [2]. To accomplish that, weak consistency models use synchronization variables. With the synchronization variables we also establish a critical section.

### 2.2. Replication Objects

A replication object is a database object existing on all servers in a replication system. In a replication environment, any updates made to a replication object at one site are applied to all other sites. Advanced Replication enables the replication of the following types of objects: Tables, Views, Indexes, Packages, Procedures and Functions, Triggers, User-Defined Types and Type Bodies.

### 2.3. Replication Groups

A replication group is a collection of replication objects that are logically related. By organizing related database objects within a replication group, it is easier to administer many objects together. Typically, a replication group is created and used to organize the schema objects necessary to support a particular database application.

### 2.4. Types of Replication Environments

Advanced Replication supports the following types of replication environments:

    a. Multimaster Replication

    b. Materialized View Replication

    c. Multimaster and Materialized View Hybrid Configurations

### 2.4.1. Multimaster Replication

Multimaster replication enables multiple sites, acting as equal peers, to manage groups of replicated database objects. Each site in a multimaster replication environment is a master site, and each site communicates with the other master sites, figure 1.

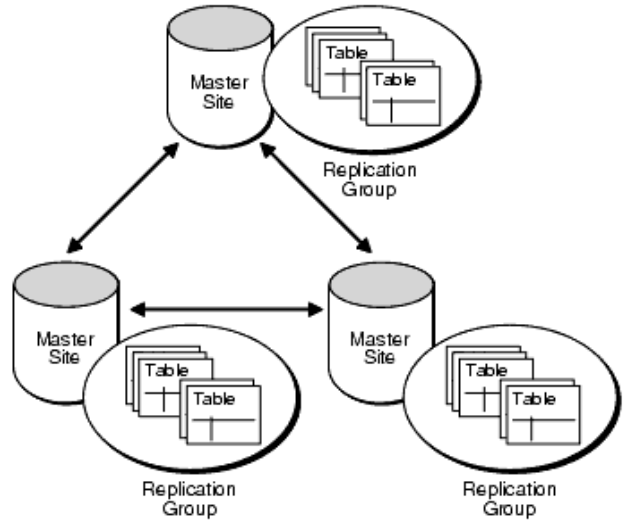

Figure 1. Multimaster replication

During administrative tasks on the master group we must stop all replication activity. For example, all replication activity for a master group should be stopped to add a new master group object. Stopping all replication activity for a master group is called quiescing the group. When a master group is quiesced, users cannot issue DML statements on any of the objects in the master group. However, users can continue to query the tables in a quiesced master group.

### 2.4.2. Materialized View Replication

A materialized view takes a different approach in which the query result is cached as a concrete table that may be updated from the original base tables from time to time. This enables much more efficient access, at the cost of some data being potentially out-of-date, figure 2. There are 3 types of Materialized Views "Read-only views", "Updatable views" and "Writable views".

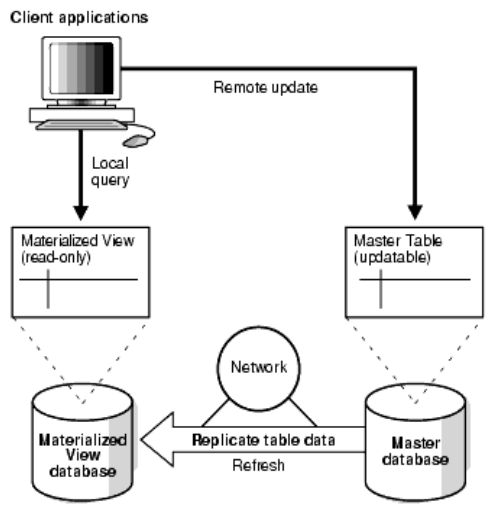

Figure 2. Materialized View

### 2.4.3. Multimaster and Materialized View Hybrid Configurations

Multimaster replication and materialized views can be combined in hybrid or "mixed" configurations to meet different application requirements. Hybrid configurations can have any number of master sites and multiple Materialized View sites for each master, figure 3.

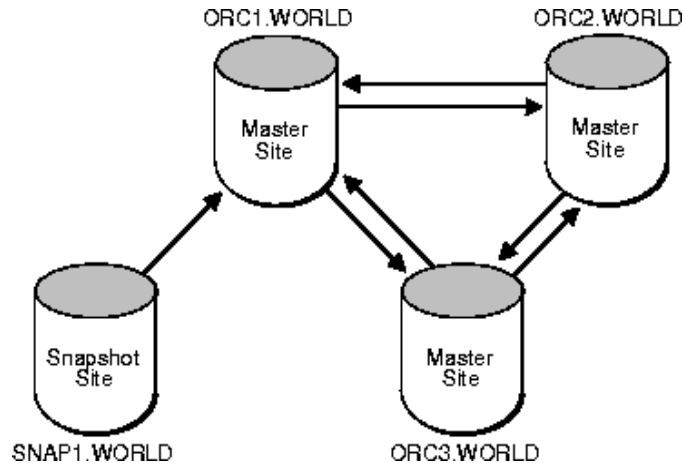

Figure 3. Hybrid Configurations

### 2.4.3. Differences between Materialized View and Multimaster

In the Materialized View Replication there is one master site so all data is stored first in the master site and then the master site forwards all DML statements to all sites. If the master site goes down for some reason, the replication activity is automatically stopped [4].

In a Materialized View Replication there is more network traffic. This problem does not exist in the Multimaster Replication.

Another problem is that in Materialized View Replication, during the data refresh from master sites, users connected to the site database cannot make queries or execute DML statements.

### 2.5. Conflicts

Conflicts only exist in asynchronous replication due to propagation to all replicas not at the same time. While some updates are carried out at a later point of time, new updates can happen. Normally for detection of reconcile updates are used timestamps, every object carries a timestamp of its most recent update. Thereby an update transaction carries the new value and also the old object timestamp with it. If this object timestamp is not different from the one of the local replica, the update is safe. If the timestamp is different, then the node may reject the incoming transaction and submits it for reconciliation [5]. There exist four types of replication conflicts:

- **Update conflicts:** An update conflict can happen when two transactions originating from different sites update the same row at the same time. This kind of conflict can be detected the by comparing the old timestamp of the replica from the propagation node with the one of the subscriber. If the timestamps differ we have an update conflict.

- **Uniqueness conflicts:** A uniqueness conflict occurs when the replication of a row attempts to violate entity integrity, such as a primary key or unique constraint.

- **Delete conflicts:** This type of conflict occurs when a transaction from one site deletes a row and another transaction from another site deletes or updates the same row, because in this case the row does not exist to be either updated or deleted.

- **Ordering conflicts:** Ordering conflicts can occur in replication environments with three or more master sites. If propagation to master site A is not allowed at a certain moment of time for any reason, then updates to replicated data can continue to be propagated among other master sites. When propagation resumes, these updates may be propagated to site A in a different order than they occurred on the other masters and these updates may conflict.

## 3. ADMINISTRATIVE TASKS IN MULTIMASTER REPLICATION

With the development of large software the probability of bugs in the software increases. In order to provide patches, updates to the objects should be prepared. It is possible that these updates should be made to replication objects bringing the need of doing some administrative tasks on the replication group. During administrative tasks the replication activity for the related replication group must be stopped. While administrative tasks are performed users cannot issue DML statements on any object of the replication group. In order to add a new object or update an existing one, 3 to 4 minutes are needed to stop the replication activity. This amount of time is not a problem because usually the interval between two data refreshes is more than 5 minutes.

### 3.1. Adding a new site in the distributed system

Adding a new site in the distributed system is another administrative task that should be considered. In this process the replication activity must be stopped, replication objects in the new site must be created, the existing data should be copied on the new database server and finally the restart of the replication can be performed. This means that during this task the system is down. The following sections explain two different ways of adding a new site.

### 3.2. Adding New Master Sites with Online Instantiation

This is the standard method of adding a new site. But for large systems this method cannot be implemented because the replication activity must be stopped for hours. During tests performed in a system with 250 tables and about 35 million records with a communication rate of 1 Mbit/s between sites, the time amount consumed by this method is too long (about 30 hours). The replication activity cannot be stopped for 30 hours on systems that should be working 24x7.

The procedure of adding a new site with Online Instantiation follows:

```
BEGIN
    DBMS_REPCAT.SUSPEND_MASTER_ACTIVITY (
        gname => 'replication_group');
END;

/* the replication will be suspended and we cannot execute DML
statements during adding new site with the following script */

BEGIN
    DBMS_REPCAT.ADD_MASTER_DATABASE (
        gname => 'replication_group',
        master => 'dbx.rep',
        use_existing_objects => FALSE,
        copy_rows => TRUE,
        propagation_mode => 'ASYNCHRONOUS');
END;

BEGIN
    DBMS_REPCAT.RESUME_MASTER_ACTIVITY
            ( gname => 'replication_group');
END;
```

## 3.3. Adding New Master Sites with Offline Instantiation

In order to minimize the time of the task of adding a new Master Site, Oracle has built another method named Offline Instantiation. The same logic applies in adding new Master Sites with Offline Instantiation as for the Online Instantiation, but this method reduces the amount of time needed by eliminating the process of copying data when the replication activity is stopped. The solution is provided by exporting the tables of the replication group which can be imported only after the replication activity is restarted. In conclusion this method eliminates the time overhead of transferring data, inserting and the building of indexes. The tests performed in the same conditions as for the Online Instantiation show a dramatic reduction of the amount of time needed (about 150 minutes). However, this amount of time is still a big problem for systems that are supposed to work 24x7.

The procedure of adding a new site with Offline Instantiation follows:

Step 1: Stopping the replication activity

```
BEGIN
   DBMS_REPCAT.SUSPEND_MASTER_ACTIVITY (
      gname => 'replication_group');
END;
```

Step 2: Beginning Instantiation

```
BEGIN
   DBMS_OFFLINE_OG.BEGIN_INSTANTIATION (
      gname => 'replication_group',
      new_site => 'dbx.rep');
END;
```

Step 3: Exporting data

```
/* from shell or command promt */
exp system/pass SCHEMAS=schema DUMPFILE=  exp.dmp
```

Step 4: Restarting replication

```
BEGIN
   DBMS_OFFLINE_OG.RESUME_SUBSET_OF_MASTERS (
       gname => 'replication_group',
       new_site => 'dbx.rep');
END;
```

Step 5: Importing data to the new site

```
BEGIN
   DBMS_OFFLINE_OG.BEGIN_LOAD (
         gname 'replication_group',new_site => 'dbx.rep');
END;

/* from shell or command promt */
imp system/pass SCHEMAS= schema DUMPFILE= exp.dmp

BEGIN
   DBMS_OFFLINE_OG.END_LOAD (
      gname 'replication_group',
      new_site => 'dbx.rep');
END;
```

Step 6: End procedure

```
BEGIN
   DBMS_OFFLINE_OG.END_INSTANTIATION (
```

```
            gname 'replication_group',
        new_site => 'dbx.rep');
END;
```

## 4. ELIMINATION OF THE TIME NEEDED FOR ADDING A NEW SITE

To eliminate the "system down" time during the adding of a new Master Site in a Multimaster Replication System we propose a new method that uses sessions and the possibility to make queries when the replication activity is stopped. We developed some PL/SQL scripts to modify SQL statements based on replication objects. Those scripts act as a layer between user and replication objects and scripts are active only during adding a new site process.

During addition a new site in an existing oracle multimaster replication system we cannot execute DML statements but we can make queries on replication tables. So if we solve this problem we can add a new site without stopping replication system.

To solve this problem we propose a new method which consists three parts as below:

- **Auto Package generation.** In this part scripts auto generate some new temporally table for every table which belongs replication objects collection. Those temporally tables contain all fields from parent tables and some new fields. We are using new fields for scripts in part two and three. During this period scripts generate package based on new tables. The generated scripts and temporally new tables will be deleted after addition a new site process.

- **Sessions Manager.** This part will be active during the addition a new site process and will analyses all users' sessions and if a user session contains DML statements or select statements oriented to a replication table which belongs to the replication group where we are generating support for its objects. DML statements and select statements are modified and will be oriented to packages and new tables which were auto generated in section one. So if a user executes a DML statement which inserts some rows to a replication tables will be executed successfully, because this statement will be modified and the new statement will not affect the replication table but the new table. In this way will be modified Update and Delete statements which are oriented to a replication table. We also have to modify every select statement oriented to a replication table. Those select statements will be oriented to the some new intelligent package which auto generated in section one. This intelligent package will return a recordset which contains all fields of the select statements. Also if during normal state (not during the addition a new site) select statements return new records the intelligent package will include all inserted new records on new temporally tables, will update them and exclude data that have been deleted if is necessary. So this package will uses replication tables and temporally tables during select statements.

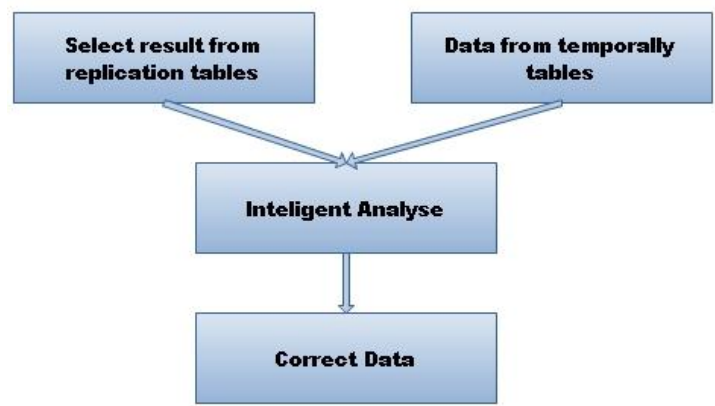

Figure 4. Sessions manager

Example: Suppose that we have a replication table as below, populated with data:

Table 1: A existing replication table

| Field_1 | Field_2 | Field_3 |
|---------|---------|---------|
| 1 | Text1 | 1234 |
| 2 | Text2 | 4321 |
| 3 | Text3 | 2233 |
| 4 | Text4 | 4411 |

During first part will auto generated a temporally table which is not populated with data as below:

Table 1: Temporally table

| Field_1 | Field_2 | Field_3 | New_1 | New_2 |
|---------|---------|---------|-------|-------|
|  |  |  |  |  |
|  |  |  |  |  |
|  |  |  |  |  |

If a user will execute an Insert statement on existing replication table we will change this statement and the new data will be inserted on temporally table. So replication table will contain only 4 existing rows and temporally table will be populated by Insert statement.

After insert statement if we execute a select statement on existing tables like "Select * from table1" result will be with 4 rows, but intelligent package will modify the select and result will be all rows on table1 and all news inserted data on temporally table.

With the same logic Update and Delete statements will be modified. So user does not understand from which table are coming those data in select statement.

- **Replication finalization.** In this session is implemented immediately after addition a new site process and will modify data on replication tables with DML statements that have been executed during the addition a new site process, those data are stored on the temporally tables. This session is implemented immediately after addition a new site process because those DML must replicate on other sites, so we cannot modify data before the replication is in a normal state. After executions of DML statements for modification of data in replication tables we have to delete all packages and temporally tables. Also we have to disable session manager so all DML statements and select statements will be automatically oriented to their original objects.

### 4.1 Ordering conflicts resolution

During adding a new site as we described above we can face some problem with ordering conflicts which means some DML statements cannot be executed but those statements are not coming from local users, they come from others replication sites.

Ordering conflicts can be resolve in a simple way, configure a job which will re execute error transactions as the receiver. To check the error queue, issue the following SELECT statement (as the replication administrator) when connected to the target master site:

*SELECT * FROM deferror;*

If the error queue contains errors, then you should resolve the error condition and re execute the deferred transaction.

*BEGIN*

```
    DBMS_DEFER_SYS.EXECUTE_ERROR ('1.12.2904', 'ORC2.WORLD');
END;
```

## 5. CONCLUSIONS

After various tests the "system down" time eliminated completely. We have implemented this method on Albanian border control replication system which includes 12 sites with more than 250 of replication tables, working 24x7 successfully.

The performance during the addition a new site process was not too high because DML statements are modified by session manager package.

Time needed to add a new site was increased but not too much because some necessary temporary tables must be generated and in finalization session are executed some DML statements.

Also during this process we must grant permissions to each user on temporary tables as the same he has on original table.

## REFERENCES

[1]   Oracle Advanced Replication, www.oracle.com

[2]   D.K. Burleson, J.Garmany, S. Karam, Oracle Replication, Rampant, 2003.

[3]   Ch. Dye, Oracle Distributed Systems, O'Reilly, 1999

[4]   "Differences between Read-Only, Updatable and Writeable Materialized Views", Metalink article, 162711.1

[5]   Bolfing F. "Database Replication with Oracle 11g and MS SQL Server 2008"

[6]   "Oracle Master Replication, Performance and Scalability", Metalink article 76448.1

[7]   M. Tumma, "Oracle Streams, High Speed Replication and Data Sharing", 2004

[8]   N. Arora, "Oracle Streams for Near Real Time Asynchronous Replication, VLDB Workshop on Design, Implementation and Deployment of Database Replication", VLDB 2005.

[9]   D. Duellmann, "Oracle Streams for the Large Hadron Collider at CERN," Oracle Open World, San Francisco, November, 2007.

[10]  Oracle Database Advanced Replication 11g Release 1 (11.1), Conflict Resolution Concepts and Architecture, www.oracle.com

**Authors**

**Hakik PACI**: PHD Student, Polytechnic University of Tirana (2011). I have a strong background in Relational Database Systems, Database Administration, and Performance improvement and in Software Design,